\documentclass[pre, twocolumn, amsmath, amssymb, superscriptaddress]{revtex4}

\usepackage{graphicx}
\usepackage{dcolumn}
\usepackage{bm}
\usepackage{braket}
\usepackage{xcolor}

\usepackage{amsmath}

\newcommand{\beq}{\begin{equation}}
\newcommand{\eeq}{\end{equation}}

\begin{document}

\title{
Robust assessment of asymmetric division in colon cancer cells
}

\author{Domenico Caudo}
\affiliation{Center for Life Nano \& Neuro Science, Istituto Italiano di Tecnologia,  Rome, Italy}

\affiliation{Department of Physics, Sapienza University, Rome, Italy}

\author{Chiara Giannattasio}
\affiliation{Department of Medical Biotechnology and Translational Medicine, University of Milan, Segrate, Italy}
\affiliation{Center for Life Nano \& Neuro Science, Istituto Italiano di Tecnologia, Rome, Italy}

\author{Simone Scalise}
\affiliation{Center for Life Nano \& Neuro Science, Istituto Italiano di Tecnologia,  Rome, Italy}

\affiliation{Department of Physics, Sapienza University, Rome, Italy}

\author{Valeria de Turris}
\affiliation{Center for Life Nano \& Neuro Science, Istituto Italiano di Tecnologia,  Rome, Italy}

\author{Fabio Giavazzi}
\affiliation{Department of Medical Biotechnology and Translational Medicine, University of Milan, Segrate, Italy}

\author{Giancarlo Ruocco}
\affiliation{Center for Life Nano \& Neuro Science, Istituto Italiano di Tecnologia,  Rome, Italy}

\affiliation{Department of Physics, Sapienza University, Rome, Italy}

\author{Giorgio Gosti}
\affiliation{Istituto di Scienze del Patrimonio Culturale, Consiglio Nazionale delle Ricerche, Montelibretti, Italy}

\author{Giovanna Peruzzi }
\affiliation{Center for Life Nano \& Neuro Science, Istituto Italiano di Tecnologia,  Rome, Italy}

\author{Mattia Miotto  \footnote{\label{corr} For correspondence write to: mattia.miotto@roma1.infn.it}}
\affiliation{Center for Life Nano \& Neuro Science, Istituto Italiano di Tecnologia,  Rome, Italy}

\begin{abstract}
Asymmetric partition of fate determinants during cell division is a hallmark of cell differentiation. Recent work suggested that such a mechanism is hijacked by cancer cells to increase both their phenotypic heterogeneity and plasticity and, in turn, their fitness.  To quantify fluctuations in the partitioning of cellular elements, imaging-based approaches are used, whose accuracy is limited by the difficulty of detecting cell divisions. 
Our work addresses this gap, proposing a general method based on high-throughput flow cytometry measurements coupled with a theoretical framework.
We applied our method to a panel of both normal and cancerous human colon cells, showing that different kinds of colon adenocarcinoma cells display very distinct extents of fluctuations in their cytoplasm partition, explained by an asymmetric division of their size.
To test the accuracy of our population-level protocol, we directly measure the inherited fractions of cellular elements from extensive time-lapses of live-cell laser scanning microscopy, finding excellent agreement across the cell types. 
Ultimately, our flow cytometry-based method promises to be accurate and easily applicable to a wide range of biological systems where the quantification of partition fluctuations would help account for the observed phenotypic heterogeneity and plasticity.
\end{abstract}

.
\maketitle

\section{Introduction}
Asymmetric cell division refers to the mechanism by which a mother cell splits into two daughter cells with distinct cellular fates. As first shown by Edwin Conklin~\cite{Conklin1905}, this process is usually achieved by asymmetric inheritance of intrinsic determinants of cell fate, such as specific proteins or RNA, and plays a crucial role during the development of organisms, favoring cell differentiation and self-renewal~\cite{Knoblich2001, Sunchu2020}.  

Increasing evidence is vouching for a more ubiquitous presence of fluctuations in the partitioning of cellular elements. In fact, asymmetric segregation has been observed in non-differentiating cell populations as different as bacteria~\cite{Mushnikov2019}, yeast~\cite{Yang2015}, and tumor cells~\cite{Buss2024, Chao2024}.
In bacterial populations, differential segregation of cellular components has been linked to antibiotic resistance~\cite{Lindner2008}.
Similarly, Katajisto \textit{et al.}~\cite{Katajisto2015}, studying mitochondrial partition in stem-like mammalian epithelial cells, found that asymmetric segregation of older mitochondria enables cells to protect themselves from aging by preventing the accumulation of misfolded proteins.
In addition, asymmetric centrosome partition has been observed in stem cells from human neuroblastoma and colorectal cancer, controlled by polarity factors and influenced by the segregation of subcellular vesicles during cell division~\cite{Izumi2012}.
To quantify the strength of fluctuations at division, the common route goes through fluorescence microscopy measurements of either fixed or live cells, where fluorescent dyes are used as a proxy for the cellular element of interest. The partition statistics are then estimated by looking at the fraction of fluorescence intensity inherited by the two daughter cells~\cite{DeyGuha2011, DeyGuha2015}.  This approach is, in principle, very informative and a lot of work has been done in recent years to write theoretical models of stochastic gene expression \cite{beentjes2020exact, jia2023coupling, dessalles2020models, thomas2018sources} that can account for the coupling of noise sources and can even be extended to time-lapse microscopy data of single cell's growing dynamics \cite{jia2021frequency}; however, relying on microscopy data implies the need of the (often manual) identification of hundreds of division events and in some cases even complex experimental set-ups and techniques where cells (usually bacteria) have to be followed for a great number of generations \cite{jia2021frequency}. These are highly time-consuming procedures that dampen the possibility of a wide characterization of the partition statistics of different cell types. \newline
To cope with these limitations,  we propose using high-throughput flow cytometry measurements to quantify fluctuations in the partitioning of cellular components in adherent cells. 
In particular, flow cytometry has already been applied to study asymmetric division in specific cellular systems: Yang \textit{et al.}~\cite{Yang2015}, sorted budding yeast cells to analyze the differential segregation of proteins between daughters cells, while Peruzzi and coworkers first applied multi-color flow cytometry to show that leukemia cells divide mitochondria and membrane proteins asymmetrically~\cite{peruzzi2021asymmetric}. Here, we generalized the protocol to adherent cell types and probed its accuracy by comparing the measured fluctuations to those observed with extensive time-lapse microscopy experiments on a panel of normal and cancerous epithelial cells. 
 \newline
Our work shows that (i) there is an exact analytical expression linking the inherited cellular component distribution dynamic to the specifics of the partition process. Based on this result, we (ii) proposed an experimental protocol to tag with fluorescent dyes cellular components and measure the dynamics of fluorescent distribution, whose analysis allows accurate estimates of the degree of asymmetry in the partition process. Applying our method to a panel of normal and cancerous human colon cells, we found that (iii) different lines of colon adenocarcinoma display very distinct extents of fluctuations in cytoplasm partition, reflecting in (iv) different degrees of asymmetric division of their size.

\begin{figure*}[]
\centering
\includegraphics[width = 0.95\textwidth]{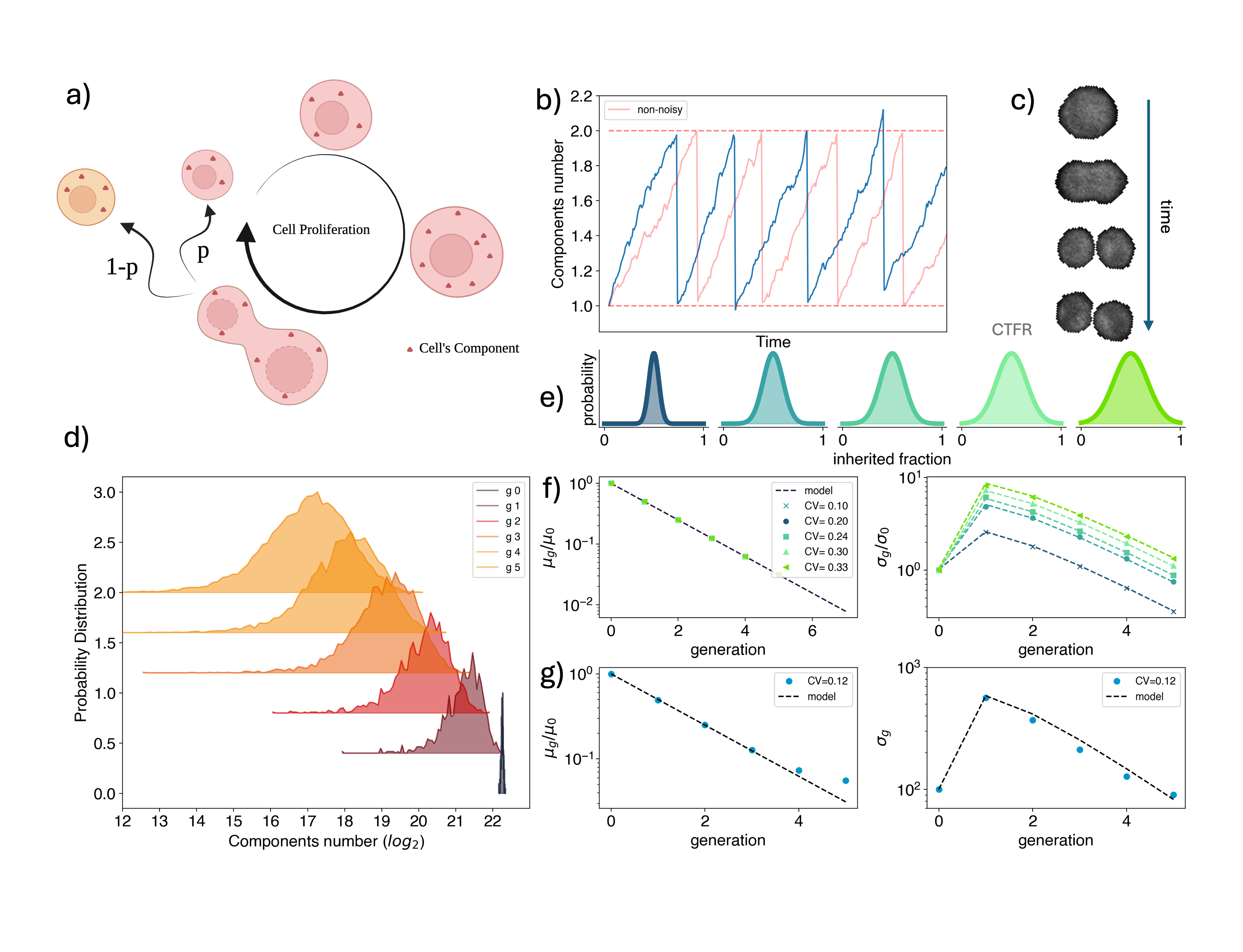}
\caption{\textbf{Modeling partitioning noise at cell division.} 
\textbf{a)} Schematic representation of the growth and division process of a mother cell. The cell undergoes an initial phase of duplication, where its internal elements are multiplied, followed by a division phase, where these elements are partitioned between the two daughter cells. One daughter inherits a fraction $p$ of the mother's elements, while the other receives the remaining fraction $q = 1-p$.  
\textbf{b)} (blue) Idealized behavior of the components number of a cell element over time, considering three different noise terms: fluctuations in compound counts during growth, uncertainty in the timing of division, and noise in the compound’s partition fraction. (red) Same description, but considering only production and degradation processes.
\textbf{c)} Microscopy images of a single cell division event for an HCT116 cell, whose cytoplasm is stained with Celltrace Yellow. 
\textbf{d)} Time evolution of the distribution of the components' number of a cellular element for a population of cells subjected only to partitioning noise. The distribution used for the simulation is a Gaussian distribution with mean $\mu = \frac{1}{2}$ and standard deviation $\sigma = 0.07$.
\textbf{e)} Examples of partition distributions $\Pi$, with increasing coefficient of variation. 
\textbf{f)} Mean ($\mu_g$) and variance ($\sigma_g$) of the number of components as a function of the generations, g, for the proliferation of a population subject to different partition noise distributions. Different colors correspond to distributions with varying coefficients of variation, as represented in the top panels. The dashed line represents the theoretical behavior obtained from the model. 
Dots are colored according to the distributions shown in panel e). 
\textbf{g)} Behavior of the mean (left) and standard deviation (right) of the simulated dynamics compared to the expected theoretical behavior for a proliferating population assuming a sizer division strategy (see SI). 
}
\label{fig:1}
\end{figure*}

\section{Results}

\subsection{Modeling cell population dynamics in the presence of partitioning noise}
\label{subsec: General Method}


In this section, we propose a theoretical framework that is able to describe the evolution of the fluorescence intensity of a cell population stained with a live fluorescent marker, by uncovering its dependence on the underlying partitioning process to which single cells are subjected.
The rationale is that, if cells follow a common division rule, knowing the shape of the mother distribution, it should be able to determine the daughters' one. By iterating this computation, one can trace the whole proliferation. The specificity of our approach lies in its specific design for live fluorescent markers. The selected component is tagged through a staining procedure with fluorophores that persist in the cell and its daughter cells, as long as the targeted protein is not degraded.
\newline
Monitoring the time evolution of the abundance of specific cellular elements over multiple divisions in single-cell experiments, it is observed~\cite{Soltani2016} that the counts of a component have a qualitatively similar dynamic to those shown in Figure~\ref{fig:1}(b). 
Indeed,  variations in the number of components are determined by: (i) the production and degradation processes along the duration of the cell growth phase; and (ii) non-identical apportioning of the components between the two daughter cells. 
Here, we will assume that the number of components varies only due to partition events and look for a quantitative description of the evolution of the element distribution. By neglecting variability in division times and the intercurrent production and degradation of components, we are reducing the dynamics to a progressive dilution. The stability of our model with respect to these assumptions will be studied in the following section and more in detail in the Supplementary Information.  Note that it is possible to model the complete growth and division process (see, for instance, refs. ~\cite{expexp, miotto2023determining, Scalise2024, jia2023coupling, thomas2018sources}), but it is more difficult to separate the different sources of noise.
Here, we start by considering a population of cells characterized by a certain initial distribution, $P(m)_0$, of element m. 
Each cell $i$ of the population divides into two daughter cells that we can label unambiguously as $2i$ and $2i + 1$, which inherit respectively  $m_{2i}$ and $m_{2i+1}$ components. The process, exactly at division, must conserve the total number of elements, i.e. $m_i = m_{2i} + m_{2i+1}$, and we call the partitioning fraction the quantity $f = \frac{m_{2i}}{m_i}$. 
To model noise in partitioning, we assume that at each division a random value for $f$ is extracted from a probability distribution function, $\Pi(f)$. Given these assumptions, we can write the probability distribution of the number of components of a daughter cell $m$ after $g$ divisions from the mother $M$ as:
\begin{equation}
    P_g(m) = \left(\frac{1}{2^{g-1}}\right)\sum_{k=0}^{2^{g-1}} \int dM P(m| M)P^k_{g-1}(M)
\end{equation}

where $P^k_{g-1}(M)$ are the components subdistributions present at generation $g-1$. After $g$ generations, the cell lineage generated from a single cell forms a lineage tree composed of $n_g = 2^g$ cells. Without losing generality, the division probability can be expressed as:
\begin{align}
    P(m|M) = \int & \delta( m - fM)\Pi(f) df~.  
    \label{eq: delta prob}
\end{align}
Therefore, we can write the probability distribution, $P(m)$,  of the daughter cells inheriting a fraction $f$ of the mother elements: 
\begin{align}
    &P(m) = \int df dM \delta( m - fM)\Pi(f) P(M) 
\end{align}
As is often the case, it is convenient to characterize the distributions in terms of their moments~\cite{miotto2023determining}. We identify with  $m_{2i}$ ($m_{2i+1}$)  the subpopulation of cells inheriting a fraction $f$ ($1-f$). In particular, the mean of the daughter cell component distribution can be expressed, for $2i$, as : 
\begin{align}
    &\mu_{2i} = E[m_{2i}] = \int P(m_{2i}) m_{2i} \, dm_{2i} \nonumber \\
    &= \int df \, dm_i \, dm_{2i} \, m_{2i} \delta(m_{2i} - fm_i) \Pi(f) P(m_i) \nonumber \\
    &= E[f] E[m_i] = \mu_f \mu_i \nonumber \\
        \label{eq: mu_mean}
\end{align}
Where $E[\cdot]$ stands for the average. Note that every computation can be symmetrically done for the siblings subpopulation $2i+1$. Thus, we have that after a single generation,  the mean number of inherited elements of the daughter cells depends on the asymmetry of the $\Pi(f)$ and on the mean of the $P(m_i)$. 
\newline
In Figure~\ref{fig:1}(d), we show the evolution of the components number distribution at different generations. The $\Pi(f)$ chosen for the simulation is a Gaussian
distribution with mean $\mu = \frac{1}{2}$ and standard deviation $\sigma = 0.7$. 
\newline
The total distribution is the sum of $2^g$ sub-populations, which is derived from the division process. For example, if we assume $\Pi(f) = \frac{1}{2}(\delta(f)+\delta(1-f))$, after one division we would have a population distributed around $f\mu$ and one at $(1-f)\mu$. After two divisions, the populations would grow to four sub-populations centered in $f^2\mu$, $f(1-f)\mu$, $(1-f)f\mu$, and $(1-f)^2\mu$, and so on. 

To compute the mean relative to an entire generation, we need to sum over all the different sub-populations: 
\begin{equation}
    \mu_g = \frac{1}{2^g} \sum_{k=1}^ {2^g} \mu_g^k
    \label{eq: mug}
\end{equation}
where $\mu_g^k$ is the mean of the k-th sub-population at generation g and $2^g$ is the total number of sub-populations for that generation. 
Equation~\ref{eq: mug} can be simplified into the form (see Supplementary Note for the full computation): 
\begin{equation}
    \mu_g = \left( \frac{1}{2} \right) ^g \mu_0
    \label{eq: mug  mu0}
\end{equation}
where $\mu_0$ is the mean of the initial population.  
It is relevant to notice that independently of the characteristic of $\Pi(f)$, the mean always halves at each generation. This result is simply understood considering that for each cell that inherits a fraction $f'$ of the compound, its sibling inherits a fraction $1-f'$. Hence, at every division, the compound count is diluted by an average factor of 2 in the population, and partition distributions with different properties lead to the same behavior of $\mu_g$ (see Figure~\ref{fig:1}(f)). Therefore, no relevant insights on the $\Pi(f)$ can be obtained from the first moment. We thus moved to consider the second moment, i.e., the variance.
Similarly to what is done for the mean, the variance of the inherited fraction of elements can be expressed as:
\begin{align}
    &\sigma_{2i}^2 = E[m_{2i}^2] - E[m_{2i}]^2 \nonumber \\
    &= \int df \, dm_i \, dm_{2i} \, m_{2i}^2 \delta(m_{2i} - fm_i) \Pi(f) P(m_i) - \mu_f^2 \mu_i^2 \nonumber \\
    &= E[f^2] E[m_i^2] - \mu_f^2 \mu_i^2 = E[f^2] \sigma_i^2 + \mu_i^2 \sigma_f^2
    \label{eq: mui mean and variance}
\end{align}
and in turn, the variance of a generation, starting from the consideration of the variance of a mixture of distributions, can be written as (see Supplementary materials for details): 
\begin{align}
    &\sigma_g^2 = \frac{1}{2^g}\sum_{k = 1}^{2^{g}} \left( \sigma_{g, k}^2 + \mu_{g,k}^2\right)-\mu_g^2 = A_g - \mu^2_g   
\label{eq: split sigmag}
\end{align}
with 
\begin{align}
    A_g = \frac{1}{2^g}\sum_{k = 1}^{2^{g}} \left( \sigma_{g, k}^2 + \mu_{g,k}^2\right)
\end{align}

It is possible to obtain a recursive equation for $A_g$ (see Supplementary Notes),  
$ A_g = A_{g-1}E[f^2] = A_0E[f^2]^g $
which leads to a compact form for Equation~\ref{eq: split sigmag}: 
\begin{equation}
    \sigma_g^2 =  \mu_0^2( E[f^2]^g - (1/2)^{2g}) + \sigma_0^2E[f^2]^g  
    \label{eq: sigmag}
\end{equation}
where $ E[f^2] = \sigma_f^2 + \mu_f^2 = \sigma_f^2 + 1/4$. 
\newline
Finally, with Equation \ref{eq: sigmag}, we can link the variance of the components distribution for the entire cell population at generation g, to the properties of $\Pi(f)$. It is interesting to highlight that there is no dependence on the bias asymmetry of the process. The distribution $\Pi(f)$ is necessarily symmetric, as daughter cells are indistinguishable. Therefore, by bias we refer to a situation where the most probable outcome of the division process is asymmetric. To this order, the only relevant feature shaping the dynamics is the extent of the fluctuations, regardless of their origin. Indeed, as shown in Figure \ref{fig:1}(e), the same $\Pi(f)$, but with a different variance, shows a different behavior for $\sigma_g^2$. \newline
Our analytical results show that the dynamical behavior of the population variance depends non-trivially on the second moment of the underlying partitioning distribution. The next section will be devoted to demonstrating how this condition can be exploited in experiments to measure the partition noise component and to evaluate the stability of our assumptions with respect to the concurrence of the other noise sources (see also Figure~\ref{fig:1}g).

\subsection{Measuring fluctuations via flow cytometry}
\label{subsec: Flow Cytometry}
As shown by Peruzzi \textit{et al.}~\cite{peruzzi2021asymmetric} for leukemia cells growing in suspension, it is possible to use flow cytometry to follow the evolution of properly marked cellular elements in time. Here, we generalized the protocol considering a panel of colon cell lines, \textit{i.e.} Caco2, a human epithelial colorectal adenocarcinoma cell line; HCT116, a human colorectal carcinoma cell line; and CCD-18Co, a normal human colon fibroblast cell line.
To measure the degree of asymmetric division, we marked cells' cytoplasm via a fluorescent dye and followed its dilution through successive generations. As explained in detail in the Materials and Methods section, the experimental procedure we used is based on the following main steps: staining of the selected compound, sorting of the initial population, plating, and acquisition of the distinct populations in the following days. The sorting allows us to obtain an initial population with a narrow peak at a specific fluorescence intensity, and it plays a dual role (a extended discussion can be found in the Supplementary Information): it indirectly induces cell synchronization and enables a clearer reconstruction of the coexisting generations.

\begin{figure*}[t]
\centering
\includegraphics[width = 0.95\textwidth]{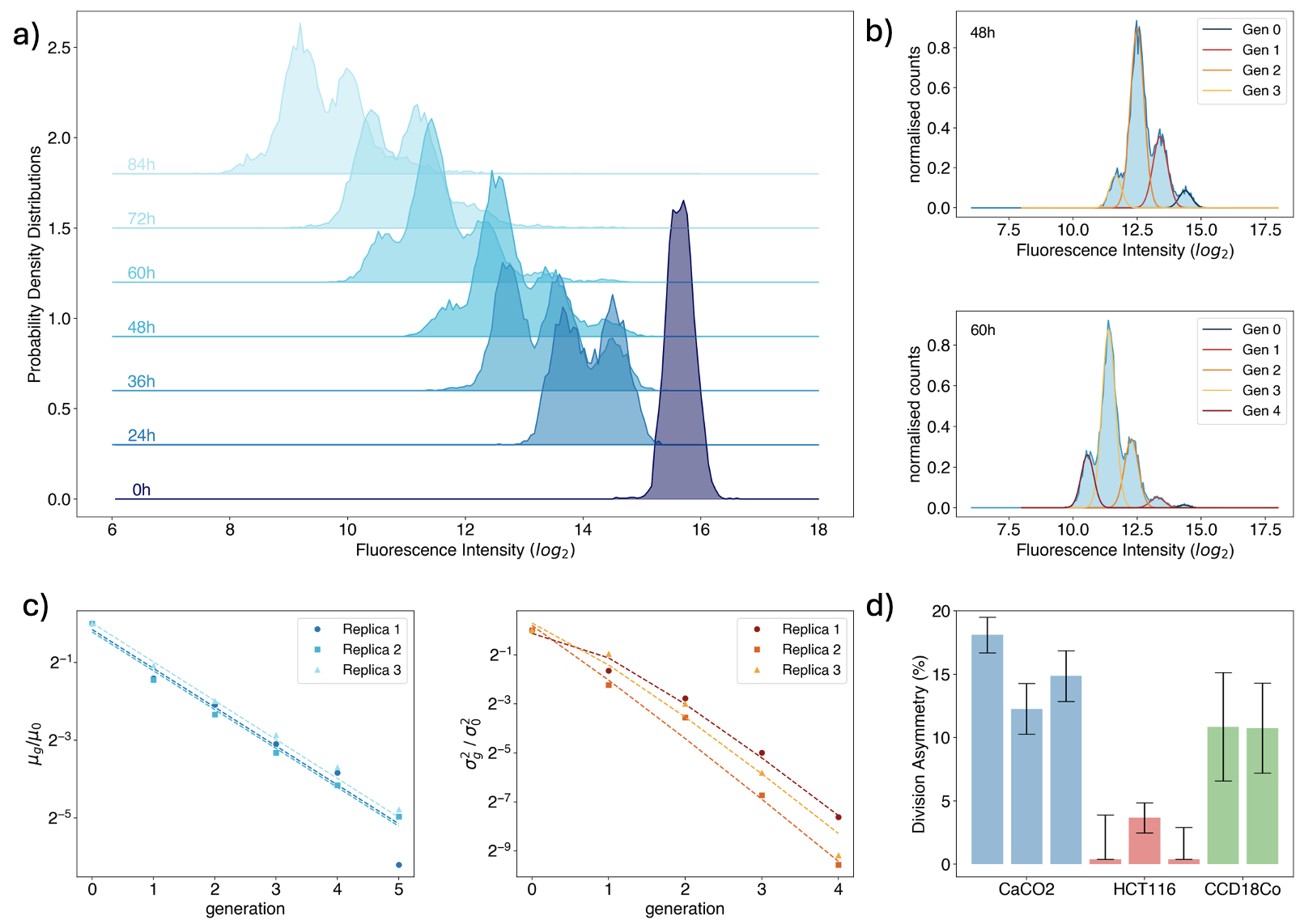}
\caption{\textbf{Quantification of partitioning noise via population-level measurements.} 
\textbf{a)} Time evolution of CellTrace-Violet\textsuperscript{TM} fluorescence intensity distribution measured in a flow cytometry time course experiment for a population of HCT116 cells. Time progresses from the darkest shade of blue to the lightest, spanning from $[0, 84]$ hours (bottom line to top). 
\textbf{b)} Snapshots of the evolution of the distribution of CellTrace-Violet fluorescence intensity measured in a flow cytometry time course experiment for a population of HCT116 cells. Experimental data are represented by the light blue histogram, while the best-fit Gaussian Mixture Model is displayed as lines, with different colors representing different generations.
\textbf{c)} Mean (left) and variance (right) of the intensity of the fluorescent markers as a function of generations, normalized to the initial population values. Each replica of the experiment is identified by a different color and a point marker's shape. The experiments are conducted on Caco2 cells, and the points correspond to the mean values for each generation. Dashed lines represent the best fit according to Equations \ref{eq: sigmag} and \ref{eq: mug mu0}, respectively.
\textbf{d)} Division asymmetry of the $\Pi(f)$ obtained by fitting Equation \ref{eq: sigmag} to data for all experiments and cell lines. The division asymmetry is measured via the percentage of the coefficient of variation.
}
\label{fig:2}
\end{figure*}

To achieve this, we used fluorescence-activated cell sorting (FACS) to isolate a specific cell population based on cell morphology and fluorescence intensity. 
The sorted population is divided into distinct wells, one per acquisition, in equal numbers and kept under identical growth conditions to mitigate inoculum-dependent variability among samples ~\cite{EnricoBena2021}.
Figure \ref{fig:2}(a) shows an example of the time course dynamics obtained with HCT116 cells (see Methods for details). Each distribution represents a different acquisition of a whole well at increasing times from plating (0 to 84 hours from the bottom to the top). 
It is possible to neatly observe the succession of peaks associated with different generations that succeed one another and the average decrease of the fluorescence intensity. Note that different generations can coexist at the same time point. One can spot the flow of cells towards higher generations and the behavior of the peak heights, which grow over time, reach dominance, and then decay. The shift on the y-axis allows us to follow the evolutionary dynamics while being able to see the overlapping of the same generation at different time points. 
As in each measurement, the distribution of fluorescent intensity, $I$, is a mixture of distributions corresponding to different generations:
\begin{equation}
    p(I) = \sum_g \pi_g P(I_g)
    \label{eq: p(I)}
\end{equation}
To obtain the necessary information on the properties of each generation, we performed a fitting of the data via a Gaussian Mixture Model (GMM) combined with an Expectation Maximization algorithm (details are explained in Materials and Methods). Indeed, assuming that each generation is log-normally distributed with mean $\mu_g$ and variance $\sigma_g^2$ of the corresponding Gaussian distribution, Equation \ref{eq: p(I)} can be rewritten as : 
\begin{equation*}
    p(log_2(I)) = \sum_g \pi_g \mathcal{N}(log_2(I_g)|\mu_g, \sigma_g) .
\end{equation*}
An example of the outcome of the GMM algorithm is shown in Figure \ref{fig:2}(b) for two acquisitions (48h and 60 h from plating). Blue-shaded distributions represent experimental data, while the best solution of the Gaussian mixture is shown with a solid line. Each generation's fit is identified with a different color. Upon an overnight, we can identify generations 0, 1, and 2 in both acquisitions. One can see that their relative fractions changed:  generation 3 became dominant overnight, and generation 4 appeared.

Figure~\ref{fig:2}(c) shows the behaviors of $\mu_g$ and $\sigma_g$ as functions of the generation number $g$ for three independent replicas of the time course experiments with CaCo2 cells. Dashed lines represent the best fit of the data against Equation \ref{eq: mug} for the mean and Equation  \ref{eq: sigmag}, which instead yields the variance $\sigma_f$ of the partition distribution probability, $\Pi(f)$. 
The outcomes of the fitting procedure on the time course dynamic for all the replicas of the three studied cell lines are reported in Figure \ref{fig:2}(d). To evaluate the degree of asymmetry, we have defined the \textit{division asymmetry} as the percentage of the coefficient of variation of the obtained partition distribution (CV: $\frac{\sigma}{\mu} $). In particular, Caco2 and CCD18Co cells show a higher level of asymmetry with respect to HCT116.  
As already stated in the description of the theoretical model, we consider only partitioning noise and neglect other potential sources: (i) production and degradation processes, and (ii) variability in cell cycle length. Since we use non-endogenous live fluorescent dyes, production processes are excluded by definition, while degradation results in a negligible time-dependent decrease in mean fluorescence intensity as the used dyes are optimized to maintain stable fluorescence levels for the duration of the experiment.  
Indeed, by looking at the data in Figure \ref{fig:2} (c), the values of the measured $\mu_g$ seem to only fluctuate around the theoretical line, but with the same slope, indicating that degradation can be neglected without major consequences.
The effects of correlations between the inherited fractions of cellular components and the fluctuations in the duration of the cell cycle are, in principle, more difficult to disentangle. In this respect, it is widely known that cells adjust their growth/division strategy to mitigate size fluctuations at division \cite{miotto2024size, thomas2018analysis}. According to the `sizer' model, they divide upon reaching a certain size so that smaller cells, i.e., those that inherit a smaller fraction of their mother’s volume, will have longer division times and vice versa. This results in fluctuations in the cell division times that may be coupled with partitioning noise. 

To assess the robustness of our model under such conditions, we performed simulations assuming that cells follow a sizer-like division strategy, while the marked component partition perfectly matches the cell size partition (see Supplementary Information for more details). This scenario maximizes the coupling between the two noise sources. As can be observed in Figure \ref{fig:1}(g), the behaviors of the mean value $\mu_g$ and the variance $\sigma_g$ deviate from the case in which division time is uncoupled with partition noise. However, the values obtained are still within the estimated experimental error (see Supplementary Figure 3). 
It is worth noticing that the presence/absence of the increase in the mean, $\mu_g$, for newer generations may highlight the presence of the coupling between the two noise sources.

\subsection{Validation of the method via live fluorescence microscopy}
\begin{figure*}[t]
\centering
\includegraphics[width = \textwidth]{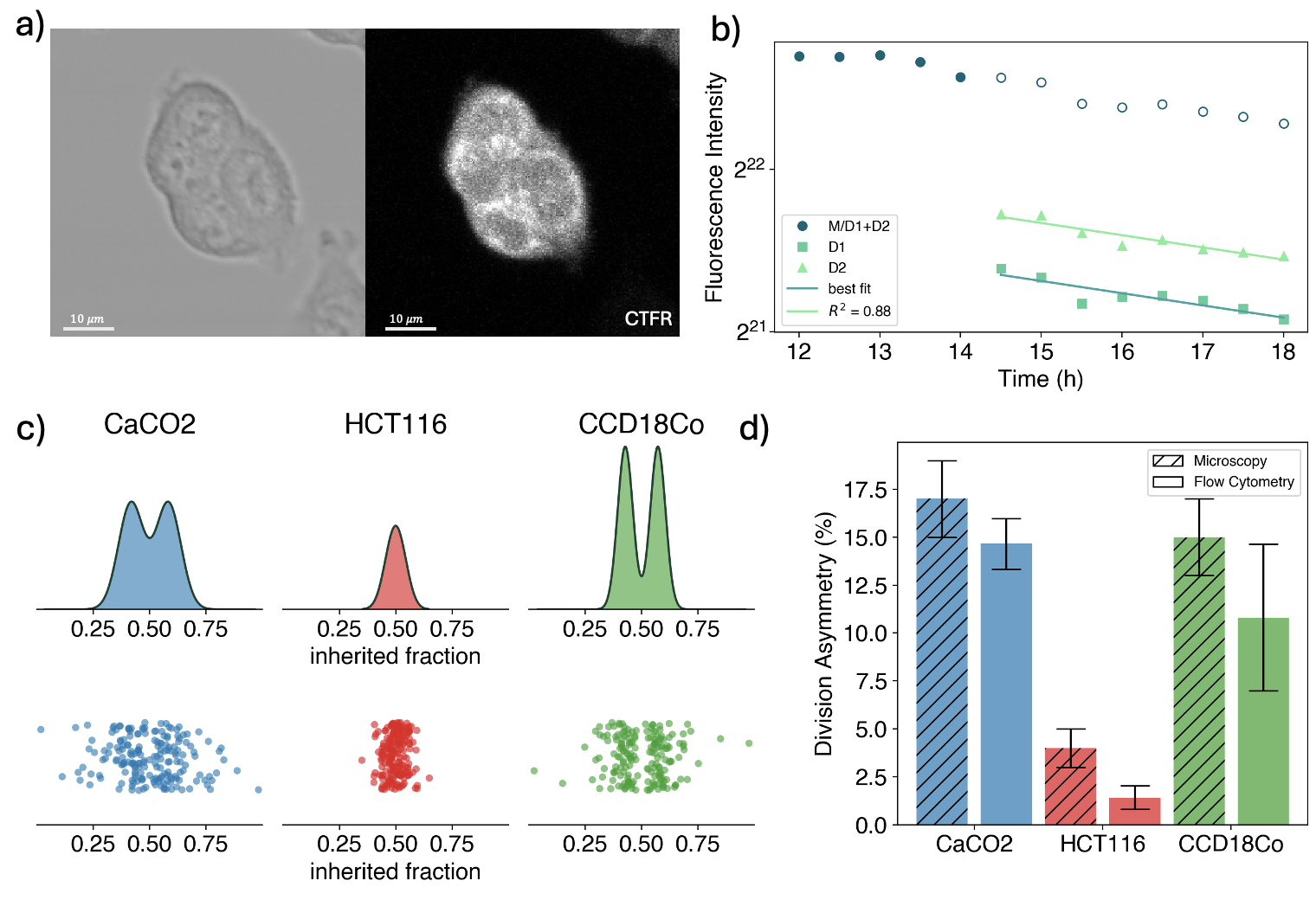}
\caption{\textbf{Quantification of partitioning noise via single-cell measurements.}
\textbf{a)} Example of a recorded cell colony of HCT116 cells in brightfield (left) and on CTFR fluorescence (right).
\textbf{b)} Cell cytoplasm fluorescence intensity as a function of time for a cell before and after division. Dark green circles correspond to the fluorescence intensity of the mother cell up to the division frame, and then to the sum of the daughters' fluorescence. Lighter green triangles and squares represent the fluorescence intensity of the daughter cells. Solid lines are the linear fit of the points.
The intercepts of the linear fit are used to compute the fraction of tagged components inherited by the daughter cells. The time is counted from the start of the experiment. 
\textbf{c)} (bottom) Strip plot of the distribution of the inherited fraction of cytoplasm for the different cell lines. The points are randomly spread on the y-axis to avoid overlay. (top) Fit of the inherited fraction distribution with the sum of two Gaussians with a mean symmetric to $1/2$.  
\textbf{d)}  Comparison of division asymmetry obtained with time-lapse fluorescent microscopy measures (striped bars) with the ones obtained from flow cytometry experiments (plain bars). The flow cytometry bars are obtained as the mean over the multiple conducted experiments. }
\label{fig:3}
\end{figure*}

The flow cytometry-based protocol combined with the developed theoretical model provides insights into the properties of the cell partition function (see Figure \ref{fig:2}). 
To validate our approach, we performed extensive live time-lapse microscopy experiments to measure the partition fraction of a labelled compound during proliferation, providing a direct measure of the $\Pi(f)$. 
Specifically, we aimed to follow a growing cell colony over time, detect divisions, evaluate the fluorescence intensity of mother and daughter cells, and calculate the inherited fraction.
The detailed experimental procedure is described in the Materials and Methods. Briefly, cells are stained for the cytoplasm and are plated at low density (approximately $10^3$  cells/well) on IBIDI cell imaging chambers ($\mu$-Slide 4 and 8 wells). After 24 hours (sufficient time for complete cell adhesion to the slide), the wells are washed, the media is renewed, and live imaging acquisition time-lapse is started. Multiple fields are followed in each well, and for each field, we acquire a z-stack bright-field and fluorescence image (a single plane is shown in Figure \ref{fig:3}a) at constant intervals (20 min) for 3 days. The fields are manually selected at the beginning of the acquisition based on cell density and fluorescence intensity. 
The recorded time lapses have been manually analyzed to identify divisions.
A sample outcome of the analysis procedure is shown in Figure \ref{fig:3}(b).
The total fluorescence of the cells is displayed as a function of time, before and after the division. The mother cell fluorescence decays exponentially over time, and at division, it halves due to components partitioning between the two daughters. In this specific case, we can observe that the fraction of cytoplasm inherited by the two cells is not equal, indicating an asymmetrical division. To compute the partition fraction $f$, we fit the fluorescence intensity of the daughters' cells vs time with $log(I_{2i}(t))  = mt + q_{2i}$ and $log(I_{21+1}(t)) = mt+q_{21+1}$, where $I(t)$ is the fluorescence intensity at time $t$, $m$ accounts for the decaying process and $2i$ and $2i+1$ respectively refer to the two daughter cells. We constrain the slope to be the same between the two cells.
The  fraction of inherited fluorescence, $f$ , is obtained as : 
\begin{equation}
    f_{2i} =  \frac{e^{q_{2i}}}{e^{q_{2i}}+e^{q_{2i+1}}} \quad \text{and}\quad
    f_{2i+1} = 1- f_{2i}
\end{equation}
To ensure the highest possible accuracy, we (i) measured the total fluorescence of the mother cell and those of the two daughter cells for at least 2 hours before and after the division event to increase the determination of the fluorescence splitting for each detected mitosis. Moreover, (ii) we removed all dynamics that showed an absolute Pearson correlation value between luminosity and pixel size higher than 0.9 as a function of time (see Supplementary Information). In fact, single cell's pixel size, which corresponds to the cell size projection on the focal plane, is found to shrink before mitosis and to enlarge after division. Fluorescent intensity is instead proportional to the number of stained cytoplasmic proteins, which are expected to remain constant (except for a constant decay of the fluorescence). Therefore, a high correlation between pixel size and fluorescence intensity indicates a high noise-to-signal ratio. 
Figure \ref{fig:3}(c)(bottom) displays the strip plot of the obtained partition fraction, $f$,  for Caco2, HCT116, and CCD18Co cells, randomly spread on the y-axis, and their corresponding fitted $\Pi(f)$ distributions (top). Note that via microscopy imaging, it is possible to measure the whole partition distribution and not just its moments; however, hundreds of events are required to get a reliable estimate. Here, we characterize it by fitting the data with a double Gaussian distribution of the form:
\begin{equation*}
    \mathcal{N}(1/2, \sigma) = \frac{\mathcal{N}(f, \sigma')+\mathcal{N}(1-f, \sigma')}{2}
\end{equation*}
that allows for the measure of both  $\langle f\rangle$ and $\sigma_f^2= \sigma'^2 + f^2 - f + 1/4$. 
\newline 
The fitted distributions clearly show how HCT116 cells are the most symmetric, with only one central and narrow peak, while for CCD18Co and CaCO2 cells, two asymmetric peaks are visible, associated with higher fluctuations.
In Figure~\ref{fig:3}(d), we compare the outcomes of the flow cytometry vs microscopy measurements. 
For all three cell lines, the found degrees of asymmetry are statistically consistent between the two approaches.

\subsection{Size division bias accounts for cytoplasmic fluctuations}

\begin{figure*}[t]
\centering
\includegraphics[width = \textwidth]{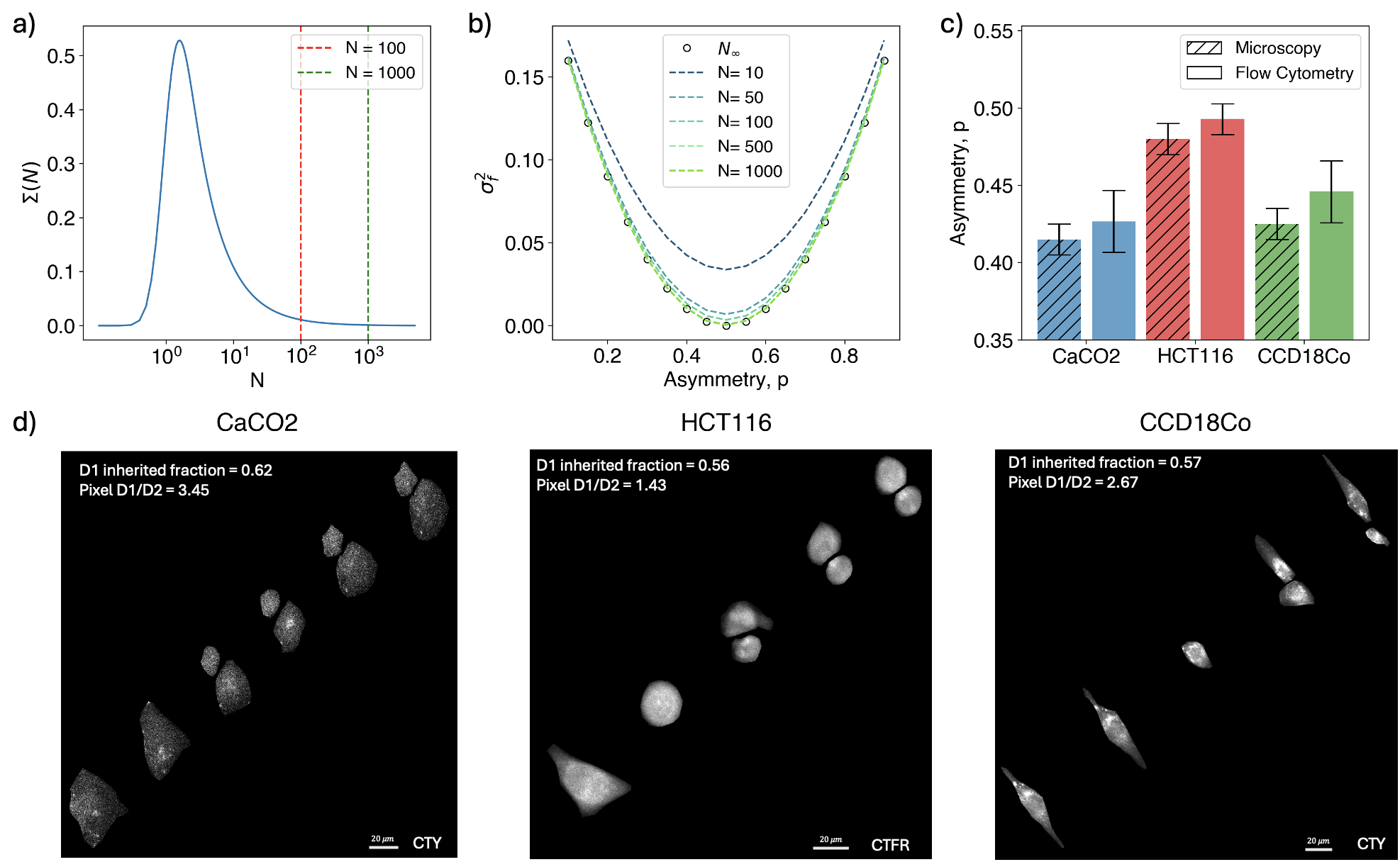}
\caption{\textbf{Cytoplasm partition fluctuations vs cells size.} 
\textbf{a)} Behavior of the integral term $\Sigma(N)$ (Eq.\ref{sigma_VERA}) as a function of the number of dividing elements, $N$,   at fixed $\sigma_{N_i} = 0.8$. Vertical lines mark typical values of cellular elements, like mitochondria. \textbf{b)} Theoretical value of the variance of $\Pi(f)$ in the binomial assumption for different levels of asymmetries and increasing values of $N$. 
\textbf{c)} Asymmetry of the partitioning distribution $\Pi(f)$ in the binomial limit as measured by the binomial bias, p.  
\textbf{d)} Sample cases of volume asymmetric division for different cell lines. Images show the overlay of consecutive times during the division dynamic. Time flows from the bottom left to the top right. 
}
\label{fig:4}
\end{figure*}
In the above sections, we have proposed and validated a method to evaluate fluctuations in the division of adherent cells. The model makes no assumptions about the shape of the partitioning distribution and, in its general form, accounts only for the effect of its variance in shaping the population dynamics. Indeed, we cannot yet distinguish between divisions which are biased, defined by a partitioning distribution with peak values different from $1/2$ (as shown in Figure \ref{fig:3}(C) for CaCO2 and CCD18Co cells) or symmetric ones (as shown in Figure \ref{fig:3}(c) for HCT-116 cells) which though have higher fluctuations. Also, apart from measuring it, we want to propose a qualitative interpretation of the obtained level of asymmetry. 
We start by recalling that our experimental protocol uses dyes that bind a-specifically to cytoplasmic amines. Thus, (i) it is expected to be uniformly distributed in the cellular cytoplasmic space and (ii) the number of labeled cytoplasmic components can be considered large. With this hypothesis, the least complex model one can assume is a binomial one with parameter $p$, measuring the bias in the process.
Via equation ~\ref{eq: sigmag}, we got a direct link between the measurable variance of the population and the second moment of the underlying partition probability distribution, $\Pi(f)$. 
Henceforth, we sought for a relationship between the second moment of $\Pi(f)$ and the parameters of the binomial distribution. 
\newline
To begin with, we can write the partition distribution for the fraction of inherited component, assuming a level of asymmetry $p$ ($q = 1-p$), as: 
\begin{equation}
    \Pi(f) = \frac{1}{2} (\Pi(f)^{(p)}+\Pi(f)^{(q)}).
    \label{distribuzione_somma}
\end{equation}
As we have already observed in the general model, due to symmetry, $\langle f\rangle = 1/2$ for any form of the $\Pi(f)$.
No information on the system can be obtained by looking at the first moment. 
\newline
For the second moment, we note that
the variance ${\sigma^2_f}^p$ of a single branch of the $\Pi(f)$ is: 
\begin{flalign}
     {\sigma^{2}_f}^{(p)}  & = {\langle f^{2}\rangle}^{(p)} - {\langle f\rangle^{2}}^{(p)}  \notag \\ 
     & = \int dN_{i} {\sigma^{2}_{f | N_{i}}}^{(p)}P(N_{i})
     \label{sigma_p_moltiN}
\end{flalign}
where $P(N_i)$ is the probability distribution of the mother’s components number at the moment of division, while ${\sigma^{2}_{f | N_{i}}}^{(p)}$, is the variance of the $\Pi(f)^p$ given the number of internal component $N_i$ in the mother cell. 
\newline 
Recalling that $m_{2i}$ is the number of inherited intracellular components of one of the two daughter cells, the partitioning fraction $f$ can be expressed as $f = m_{2i}/N_i$, which leads to: 

\begin{flalign}
    {\sigma^{2}_{f | N_{i}}}^{(p)} & = {\langle f^{2}\rangle _{N_i}}^{(p)} - {\langle f\rangle^{2}_{N_i}}^{(p)} \notag \\ 
    & = \frac{1}{N_{i}^{2}} {\sigma^{2}_{m_i}}^{(p)} = \frac{1}{N_{i}^{2}} N_{i}pq = \frac{pq}{N_{i}}.
    \label{unica_sigma}
\end{flalign}
Therefore, by substituting it into equation \ref{sigma_p_moltiN} we obtain: 
\begin{equation}
    {\sigma^{2}_f}^{(p)} = pq \int dN_i \frac{1}{N_{i}} P(N_{i}) ,
    \label{sigma_tantiN_versionefinale}
\end{equation}
Which explicitly depends on mother's component probability distribution, $P(N_i)$,  and coherently returns ${\sigma^{2}_{m_i}}^{(p)} = \frac{pq}{N}$ for $P(N_i) = \delta(N_i-N)$. 
\newline
An equivalent computation can be done for the symmetric branch ($ {\sigma^{2}_{m_i}}^{(p)}$) and therefore the variance of the full distribution is given by:
\begin{flalign}
\sigma^{2}_{f} & = \langle f^{2}\rangle - \langle f\rangle^{2}\notag \\ 
    & = \frac{1}{2}({\langle f^{2}\rangle}^{(p)} - {\langle f^{2}\rangle}^{(q)}) - \frac{1}{4} \notag \\ 
    & = pq \int dN_{i} \frac{1}{N_{i}}P(N_{i}) + \frac{p^{2}+ q^{2}}{2} - \frac{1}{4} .
\label{sigma_VERA}
\end{flalign}

Note that this relationship depends on the integral: 
\begin{equation*}
    \Sigma(N) = \int dN_{i} \frac{1}{N_{i}}P(N_{i})
\end{equation*}
which requires the knowledge of the mother element distribution $P(N_i)$ to be computed. 
To get an idea of its behavior, we observe that assuming a population of cells with a fixed number of elements before division, i.e.  $P(N_i) = \delta(N-N_i)$, the integral simply goes as $1/N$;
in a more realistic scenario, the elements can be considered  log-normally distributed, since they are the product of multiple division processes, hence: 

\begin{equation*}
     P(N_i) = \frac{1}{N_i\sqrt{2\pi\sigma_{N_{i}}^{2}}}
\text{exp}\left(\frac{-(ln(N_{i})-\mu)^{2})}{2\sigma_{N_{i}}^{2}}\right)
\end{equation*}
    
In this case, one gets the following expression for $\sigma_f^2$: 
\begin{flalign}
    \sigma^{2}_{f} 
    = \frac{1}{2} e^{\sigma_{N_i}^{2}/2-\mu}(1 - erf(\frac{\sigma_{N_i}^{2}-\mu}{\sqrt(2)\sigma_{N_i}})) + \frac{p^{2}+ q^{2}}{2} - \frac{1}{4}
    \label{eq: sigma N lognormal}
\end{flalign}
where $\mu = \langle log(N_i)\rangle$. 
\newline 
The quantity $\sigma_{N_i}$ can be computed from flow cytometry or microscopy measurements. Indeed, if we consider the relation $I = N_i F$, where $I$ is the total fluorescence intensity, $N_i$ the number of components, and $F$ the fluorescence of each component, then $\sigma_I^2 = \sigma_{N_i}^2$. Since $\sigma_I^2$ is accessible from the data, so is $\sigma_{N_i}$.  On the other hand, the values of $\mu$ are not directly accessible, limiting the possibility of an exact analytical computation. However, by computing the value of $\Sigma(N)$ as a function of $N$ for fixed values of $\sigma_{N_i}$ (Figure \ref{fig:4}(a)), one can see that already for $N = 100$ the role of $\Sigma(N)$ is negligible. 
In general,  for the partitioning of components such as cytoplasm, which are present in much larger numbers in the cell, this factor can be neglected. 
In this regime, the expression for $\sigma_f^2$ becomes: 
\begin{equation}
    \sigma^{2}_{f} = \frac{p^{2}+ q^{2}}{2} - \frac{1}{4} .
    \label{eq: sigma N inf limit}
\end{equation}
By inverting this relationship, we obtained a general mapping between the variance and the level of biased asymmetry of the binomial distribution. 
\begin{equation}
    p = \frac{1}{2} - \sigma_{f} 
    \label{eq: p vs sigma binomial}
\end{equation}
In Figure \ref{fig:4}(c), we validated the analytical expression against the values obtained via microscopy measurements, showing again good compatibility of the results.
CaCO2 cells confirm to be the most asymmetric among the cell lines, and since they are known for their heterogeneity in cell morphology~\cite{Lea2015}, we explored the hypothesis that shape heterogeneity and cytoplasmic asymmetry are linked.
Figure \ref{fig:4}(d) displays three sample cases of division examined through time-lapse fluorescent microscopy following cytoplasm. Different time frames are overlaid to depict the entire dynamic at once, with time progressing from the bottom left to the top right of each image.
\newline
We computed the pixel size ratio between the daughter cells in the last analyzed frame and compared it to the fraction of cytoplasm inherited from the mother cell. Although pixel size is only a rough proxy for cell volume, we observe that the cell inheriting a larger fraction of cytoplasm also appears larger in size. This finding suggests that asymmetry in cytoplasmic partitioning is linked to size fluctuations at birth, where cell size reflects the bias observed in the segregation process. 

\section{Discussions}

Cell division is orchestrated by hundreds of molecular interactions, which are intrinsically stochastic processes \cite{bialek2012biophysics}. The presence of such stochasticity results in the insurgence of variability among the cell phenotypes sharing the same genome, as is the case for cancer cells.
In this respect, gene expression noise has been extensively characterized \cite{Elowitz2002}, incorporating cell cycle effects \cite{thomas2019intrinsic, swain2002intrinsic, chen2004dynamics}, as well as the different strategies that cells evolved to regulate the extent of the noise in these channels \cite{Osella2011, Miotto2019mirna}. 

Besides such noise sources, proliferating cells are subject to the fluctuations originating from the partition of cellular elements at division. Analysis of this partition noise highlighted that under some circumstances, its contribution exceeds the others in creating heterogeneity \cite{Huh2011, Soltani2016}.
In fact, asymmetrically dividing cells can produce daughters that differ in size, cellular components, and, in turn, fate \cite{Chhabra2021}.

In this framework, the origin of such asymmetry may be ascribed to the basal stochasticity of molecular processes taking place at different levels during the division and/or to specific, evolutionarily conserved mechanisms used by cells to control cell fate and generate diversity \cite{Sunchu2020, evolat}. In fact, asymmetric partitioning of components alters the initial counts of molecules in the following cell cycle, which can lead the system to a different phenotypic state \cite{moris2016transition, Miotto2023coll}.
Here, we show that an accurate determination of partition fluctuations can be obtained via standard flow cytometry measurements of properly marked cell populations. Our approach is easier and faster with respect to the use of microscopy imaging, which requires the acquisition and analysis of extensive recordings of the cells, ultimately limiting the reachable statistics. In addition, it is specifically designed to isolate partitioning noise from other sources via the usage of non-endogenous dyes to target cell components.   \newline 
Although the need to use live fluorescent dyes to track proliferation surely limits the extent of applicability of our method to markable components, we anticipate that there are numerous contexts in which our approach could give a determining insight. For instance, organelles such as mitochondria, the endoplasmic reticulum, lysosomes, peroxisomes, and centrosomes can be easily marked~\cite{peruzzi2021asymmetric, Miotto2025phyl} and their unequal distribution between daughter cells has been observed to provoke functional differences that influence their fate. In this respect, asymmetric division of mitochondria in stem cells is associated with the retention of stemness traits in one daughter cell and differentiation in the other \cite{Katajisto2015} and, although the functional consequences are not yet fully determined, with therapeutic resistance in Cancer Stem Cells \cite{hitomi2021asymmetric, garcia2020role}.  Asymmetric inheritance of lysosomes,  together with mitochondria, appears to play a crucial role in determining the fate of human blood stem cells \cite{loeffler2022asymmetric}. Furthermore, our approach could be extended to investigate cell growth dynamics and the origins of cell size homeostasis in adherent cells \cite{miotto2023determining, miotto2024size,kussell2005phenotypic,mcgranahan2017clonal,de2019exploration}.

Here, we measured the segregation statistics of the cytoplasm of different colon cell types, finding that they are neatly distinct. CaCo2 cells exhibit the highest degree of cytoplasmic partition fluctuations, while HCT116 division is the one with the lowest asymmetry. It is interesting to note that normal colon fibroblasts sit in between, suggesting that the cancer-associated deregulation of the normal cell activities leads to the reactivation of asymmetric cell division in a cancer-specific manner. Indeed, CaCo2 cells are known to display a markedly heterogeneous distribution of phenotypes \cite{Lea2015}. Notably, comparing the degree of asymmetry in the division of the cytoplasm in the three different analysed cell types, we found that microscopy-derived fluctuations are systematically higher than those measured by flow cytometry. While this could be linked to the different protocols used, it seems more probable that the difference is due to a systematic overestimation of the fluctuations measured by microscopy. 
Indeed, we identified at least two possible origins of such bias: fluctuations in the fluorescence intensity measures and errors in the segmentation protocol. Regarding the first one, we demonstrated its effect through simple simulations that mimicked the microscopy analysis. While for the second, relying on manual segmentation not only reduces the data but might also introduce biased cell identification errors. Both these effects become important when data filtering is only partially allowed due to the low available statistics. A discussion on this topic is proposed in the Supplementary Information.

We have assessed the stability of our model with respect to perturbations arising from the co-occurrence of multiple noise sources. In particular, we explored the case where fluctuations at division and variability in cell cycle length are dependent. In particular,  we investigated a strongly coupled scenario in which division is triggered upon reaching a specific size (sizer strategy) and partitioning fluctuations correspond to size inheritance. We observed that -even under this extreme condition- our approach yields reliable parameter estimates that remain consistent with the experimental error. Moreover, we observed coupling-dependent increase in $\mu_g$ for newer generations. This observation opens new questions and potential applications for our method. \newline

Finally, to get insights into the possible mechanisms behind the observed asymmetries, we looked for a specific shape of the partition distribution. Following a maximum entropy approach \cite{bialek2012biophysics, Miotto2018, TOLOMEO}, we started by considering the distribution that employs the fewest assumptions about the system, \textit{i.e.}, we consider independent segregation, which directly maps into binomial partitioning. Indeed, this is a common assumption in segregation models that traces back to the pioneering works of Berk \cite{berg1978model} and Rigney \cite{rigney1979stochastic}. Notably, our results indicated biased segregations with colon cancer cells displaying different levels of biases. To explain their origin, we looked at the relative sizes of the daughter cells right after division. The found correlation between daughter sizes and the fractions of inherited fluorescence vouches size as the origin of the observed biased segregation. 
\newline
In conclusion, by combining experimental data with statistical modeling, we show how it is possible to use flow cytometry data to extract reliable estimates of the strength of fluctuations during cell division. Our approach has the potential to be applied to different cell types, where a quantification of the level of division asymmetries that daughter cells can experience may provide new insights into the mechanisms of asymmetric cell division and its role in cancer heterogeneity and plasticity. 

\section{Materials and Methods}
\label{sec: Materials and Methods}

\subsection{Cell Culture}
Caco-2, colorectal adenocarcinoma cell line, was purchased from ATCC (Manassas, VA, USA, HBT-37)  and maintained in complete culture media DMEM (D6046) containing $20 \%$ FBS, penicillin/streptomycin plus glutamine, nonessential amino acids (NEAA), and sodium pyruvate, all $1/100$ dilution. 
 HCT116 VIM RFP, colorectal carcinoma cell line, was purchased from ATCC (Manassas, VA, USA, CCL-247EMT) and maintained in complete culture media McCoy's 5A (M9309) containing 10$\%$ FBS,  penicillin/streptomycin, plus glutamin. \newline
 CCD-18Co, colon fibroblast cell line, was purchased from ATCC (Manassas, VA, USA, CRL-1459) and maintained in complete culture media DMEM containing $10 \%$ FBS, penicillin/streptomycin plus glutamin, nonessential amino acids (NEAA), and sodium pyruvate, all $1/100$ dilution. 
 \newline
 All cells were kept in culture at $37^\circ C$ in $5\%$ CO2 and passaged according to the experimental protocol.  For all experiments, cells were harvested,  counted, and washed twice in serum-free solutions and resuspended in room temperature (RT)  PBS w/o salts for further staining.
\newline
In detail, to detach cells from culture flasks, we used Trypsin EDTA $0.25\%$. Culture media was removed and the flasks were washed once with PBS, then Trypsin EDTA was added and allowed to work for $1$ min in an incubator at $37^\circ C$. Once the cells appeared detached, the flasks were mechanically agitated to facilitate the cells' loss of adhesion. Trypsin was then inactivated with one volume of complete media, and cells were collected, pelleted, and washed a second time in PBS. To determine cell viability, prior to dye staining, the collected cells were counted with the hemocytometer using Trypan Blue, an impermeable dye not taken up by viable cells.

\subsection{Flow Cytometry and Cell Sorting}

To track cell proliferation by dye dilution for establishing the progeny of a mother cell, cells were stained with CellTrace Violet\textsuperscript{TM} (CTV). The CTV dye staining (C34557, Life Technologies, Paisley, UK), used to monitor multiple cell generations, was performed according to the manufacturer’s instructions, diluting the CTV 1/1000 in 0.5-1 ml of PBS for 20 min in a water bath at $37^ \circ C$, mixing every 10 min. Afterward, 5x complete media was added to the cell suspension for an additional 5 min incubation before the final washing in PBS. 
\newline
Labeled cells were sorted using a FACSAriaIII (Becton Dickinson, BD Biosciences, USA) equipped with Near UV 375nm, 488nm, 561nm, and 633nm lasers and FACSDiva software (BD Biosciences version 6.1.3). Data were analyzed using FlowJo software (Tree Star, version 10.7.1). Briefly, cells were first gated on single cells, by doublet exclusion with morphology parameters area versus width  (A versus W), both side and forward scatter. The unstained sample was used to set the background fluorescence. The sorting gate was set around the maximum peak of fluorescence of the dye distribution. In this way, the collected cells were enriched for the highest fluorescence intensity. Following isolation, an aliquot of the sorted cells was analyzed with the same instrument to determine the post-sorting purity and population width, resulting in an enrichment of $\rangle$ 97 \% for each sample.
\newline
 To monitor multiple cell divisions, the sorted cell population was seeded in 12-well plates (Corning, Kennebunk, ME, USA) at a cell density between $[30-70] \cdot 10^3$ cells/well, according to the experiment,  and kept in culture for up to 84 h.  Each well corresponds to a time point of the acquisition, and cells in culture were analyzed every 24, 36, 48, 60, 72, and 84 h by the LSRFortessa flow cytometer. In order to set the time zero of the kinetic, prior to culturing, a tiny aliquot of the collected cells was analyzed immediately after sorting at the flow cytometer. The unstained sample was used to set the background fluorescence as described above.

\subsection{Time-lapse Microscopy}

To better investigate the cell proliferation dynamics, we performed time-lapse experiments for up to 3 days.
In our previous paper \cite{peruzzi2021asymmetric}, we verified that cell growth is not affected by different dye combinations. Therefore, in the present work, we used both CellTrace Yellow\textsuperscript{TM} (C34573 A, Life Technologies, Paisley, UK) and CellTrace Far Red\textsuperscript{TM} (C34572, Life Technologies, Paisley, UK) to stain the cytoplasm for microscopy analysis. 
\newline
Low passage cells at around 60 $\%$  confluency were counted and stained with CellTrace dyes 1/500 and plated on  IBIDI cell imaging chambers ($\mu$-Slide 4 and 8 wells) at a low cell density of $10^3$ cells/well. After overnight incubation, the chamber is transferred to the inverted microscope adapted with an incubator to keep cells in appropriate growing conditions.  Brightfield and fluorescent confocal image stacks were acquired with a 20x air objective (Olympus,  Shinjuku, Japan) and Zen Microscopy Software (Zeiss, Oberkochen, Germany), every 20 min.    
\newline
Time-lapse images were analyzed using ImageJ and in-house Python programs. 

\section*{Data Availability}

The data that support the findings of this study are available from the corresponding
author upon reasonable request.

\section*{Code Availability}

All codes used to produce the findings of this study are available from the corresponding author upon request. The code for the Gaussian Mixture algorithm is available at \href{https://github.com/ggosti/fcMGM}{https://github.com/ggosti/fcGMM}.\\

\section*{Author contributions statement}

M.M. and G.G. conceived the research;  D.C., G.P., and S.S. performed the cell biology experiment and flow cytometry measurements.  D.C., C.G., V.d.T., and M.M. performed timelapse microscopy experiments;
F.G. and G.R. contributed additional ideas and helped interpret results;  D.C. analyzed data; M.M. performed analytical calculations; D.C. and G.G. performed numerical simulations and statistical analysis; all authors analyzed results and wrote and revised the paper.

\section*{Competing Interests}
The authors declare no competing interests.

\section*{Acknowledgements}
This research was partially funded by grants from ERC-2019-Synergy Grant (ASTRA, n. 855923); EIC-2022-PathfinderOpen (ivBM-4PAP, n. 101098989);
Project `National Center for Gene Therapy and Drugs based on RNA Technology' (CN00000041) financed by NextGeneration EU PNRR MUR—M4C2—Action 1.4—Call `Potenziamento strutture di ricerca e creazione di campioni nazionali di R\&S' (CUP J33C22001130001).
Project FIS2023-02957 (CUP B53C24009530001), 'Fathoming oUt the role of partitioninG noIse in cancer epithelial-mesenchymal Transitions (FUGIT),  funded with the contribution of the Italian Ministry of University and Research pursuant to Ministerial Decree No. 1236 of August 1, 2023 - FIS 2 CALL.

\bibliographystyle{unsrt}
\bibliography{mybib}

\end{document}